\documentclass{aa}  

\usepackage{natbib}
\bibpunct{(}{)}{;}{a}{}{,} 

\usepackage{graphicx}
\usepackage[table] {xcolor}
\usepackage{txfonts}
\begin{document}

   \title{Limits to the presence of transiting circumbinary planets in CoRoT data}


   \author{P. Klagyivik\inst{1,2,3}\fnmsep\thanks{email: pklagyi@gmail.com}
          \and
          H. J. Deeg\inst{1,2}\fnmsep\thanks{email: hdeeg@iac.es}
          \and
          J. Cabrera\inst{4}
          \and
          Sz. Csizmadia\inst{4}
          \and
          J. M. Almenara\inst{5}
          }

   \institute{Instituto de Astrof\'\i sica de Canarias, C. Via Lactea S/N, E-38205 La Laguna, Tenerife, Spain\\
         \and
             Universidad de La Laguna, Dept. de Astrof\'\i sica, E-38206 La Laguna, Tenerife, Spain\\
         \and
             Konkoly Observatory, Research Centre for Astronomy and Earth Sciences, Hungarian Academy of Sciences, H-1121 Budapest, Konkoly Thege Mikl\'os \'ut 15-17, Hungary\\
         \and
             Institute of Planetary Research, German Aerospace Center, Rutherfordstrasse 2, D-12489, Berlin, Germany\\
         \and
             Observatoire Astronomique de l'Universit\'e de Gen\`eve, 51 chemin des Maillettes, 1290 Versoix, Switzerland
         }

   \date{}

 
  \abstract
   {}
   {The {CoRoT} mission during its flight-phase 2007-2012 delivered the light-curves for over 2000 eclipsing binaries. Data from the {\it Kepler} mission have proven the existence of several transiting circumbinary planets. Albeit light-curves from {CoRoT} have typically lower precision and shorter coverage, {CoRoT's} number of targets is similar to {\it Kepler}, and some of the known circumbinary planets could potentially be detected in {CoRoT} data as well. The aim of this work has been a revision of the entire CoRoT data-set for the presence of circumbinary planets, and the derivation of limits to the abundances of such planets.}
   {We developed a code which removes the light curve of the eclipsing binaries and searches for quasi-periodic transit-like features in a light curve after removal of binary eclipses and instrumental features. The code needs little information on the sample systems and can be used for other space missions as well, like {\it Kepler, K2, TESS and PLATO}. The code is broad in the requirements leading to detections, but was tuned to deliver an amount of detections that is manageable in a subsequent, mainly visual, revision about their nature.}
   {In the {CoRoT} sample we identified three planet candidates whose transits would have arisen from a single pass across the central binary. No candidates remained however with transit events from multiple planetary orbits. We calculated the upper limits for the number of Jupiter, Saturn and Neptune sized planets in co-planar orbits for different orbital period ranges. We found that there are much less giant planets in short-periodic orbits around close binary systems than around single stars.}
   {}


   \maketitle
%

\section{Introduction}

Several years before the detection of the first extrasolar planet, eclipsing binary systems (EBs) were already suggested as prime targets for the detection of  transiting planets \citep{1984Icar...58..121B, 1990A&A...232..251S, 1994P&SS...42..539S, 1994Ap&SS.212..335H}. This predilection for EBs came from the assumption that circumbinary planets (CBPs) - that is, planets orbiting around both binary components - would have orbital planes that are likely aligned to the binary plane. With EBs being defined as binaries for which we can observe eclipses, this means that their planets have a high chance to cause observable eclipses or transits as well. Next, the eclipsing M-star binary CM Draconis was identified as the best object for a targeted transit search \citep{1995EM&P...71..153S}, which was initiated as the 'TEP project' during the years 1994 - 2000 \citep{1996JGR...10114823D, 1996Icar..119..244J, deeg98, doyle00}. TEP remained the only search for transiting CBPs until the arrival of CoRoT, when an early part of its data were analysed for the presence of CBPs within the PhD project of J.\,M.\,\citet{almenara10}. While none of these searches found any CBPs, shortly after, the detection of Kepler-16b \citep{doyle11} proved both the existence of CBPs and the validity of transits as a method to detect them. Further detections, all in data from the Kepler mission, have presently led to a total of 9 known transiting CBPs. Including recent discoveries of CBPs by other methods, notably imaging \cite[e.g.][]{2011AJ....141..119K} and eclipse timing \cite[e.g.][]{2010A&A...521L..60B,2011MNRAS.416.2202P,2015A&A...577A.146B} a total of 22 CBPs\footnote{Given is the number of planets in the NASA Exoplanet Archive with a 'circumbinary flag' of 1, as of October 2015} are currently known and CBPs have become a very active sub-discipline of exoplanet research.

The inner edge of a stable planetary orbit around a binary system has an orbital period of $\sim$ 3 times the orbital period of the binary \citep{dvorak89,holman99}. Therefore, CBPs close to the inner stability limit in short-periodic binary systems should be more readily detectable than those around longer-period binaries. For one, the probability that such a CBP is inclined correctly for observable transits is higher than in longer-period systems, and for another, such a CBP would cause more transits in a given amount of time. These detection biases are similar to planets transiting single stars, where short-periodic planets are more readily detectable.
None of the CBPs detected by Kepler had central binaries with periods $< $ 7 days. This lack of planets orbiting short-period EBs is apparently real and not due to difficulties in their detection. \citet{hamers16} argued that short-periodic binaries formed in triple systems, followed by a dynamical evolution that either ejects their planets, or moved them to wide and potentially inclined orbits. A further large-sample search for the presence of shorter-periodic CBPs (which can only have stable orbits around correspondingly shorter-periodic binaries) is therefore of interest, as it may support the hypothesis that short-periodic CBPs are absent or very rare, and in turn may support the interpretations arising from this observation.

The {\it CoRoT} space telescope during its nearly 6 years of activity observed approximately the same number of targets as {\it Kepler}, and the number of surveyed EBs are also similar. However, there are some differences which reduce the chances of finding transits in the {\it CoRoT} eclipsing binary light curves. Its photometric noises are $\approx 4 \times$ larger than {\it Kepler's} for a given brightness (\citealt{aigrain09} for {\it CoRoT} and \citealt{2010ApJ...713L.160G}, \citealt{2010ApJ...713L.120J} for {\it Kepler}) and the time-coverage of its observations is between 30 and 180 days, depending on the observational run, in contrast to the $\approx 4$ years of {\it Kepler}. Detection probabilities in the CoRoT sample are therefore more heavily tilted towards the discovery of shorter-periodic planets than are data from Kepler, with the longest-periodic CoRoT planet around a single star being CoRoT-9b with p = 95 days \citep{2010Natur.464..384D}, whereas the longest-periodic Kepler planet is Kepler-455b with p = 1322 days \citep[][ there given as KIC 3558849 b]{2015ApJ...815..127W}.

The first search for CBPs in Corot data by \citet{almenara10} was limited to the CoRoT data that became available during 2008, namely CoRoT's 'Initial Run' (IRa01) and the first two 'Long Runs' (LRc01 and LRa01). This search was performed with a matched-filter detection algorithm based on the one by ~\citet{doyle00} used in the TEP project, which had been developed for a deep search on a single, well-characterised target. It intended a relatively detailed modelling of the CBP transit signatures, while stepping through the potential planets' orbital periods and phases.  The algorithm required therefore rather good knowledge about the binaries that are being analysed, in particular of their masses and sizes. Consequently, only a set of 10 binaries from CoRoT data were searched in detail. These had been selected by requiring detached and deep eclipses, a high S/N, and a relatively short orbital period, in order to maintain $P_{min} < T_{cov}$, with $P_{min}$ being the period of the innermost stable CBP orbit and $T_{cov}$ being CoRoT's observing-duration. A modelling of their CoRoT light-curves permitted the determination of the components' relative sizes and orbital inclinations. For some of them, additional radial velocity observations were also obtained, for a better determination of absolute and relative masses and sizes. Of several planet-like transit features detected by the algorithm, none survived on closer scrutiny. As result, only coplanar CBPs of Saturn to Jupiter sizes, with periods between the innermost stable one, $P_{min}$, and $\approx T_{cov}$ could be excluded on these few binaries.

The CoRoT mission's active phase finished in Nov. 2012, and the full data-set became available a few months later. The availability of the full CoRoT mission data, the knowledge from Kepler that transiting CBPs exist and may be detectable, and the development of an improved detection algorithm, presented in Sect.~\ref{sec:method}, provided then sufficient motivation for a renewed CBP search in the CoRoT data.

In this paper we present this method for quasi-periodic transit detection, including the necessary selection and preparation of the binary light-curves, and the results of this search for circumbinary planets in {\it CoRoT} data. In Sect.~\ref{Sec:search} we outline several issues affecting CBP transit search, with an overview of CBP detection algorithms. In Sect.~\ref{sec:bin_sample} we give a brief description of the binary sample. In Sect.~\ref{sec:method} we detail the method of the search and the code. Test results on artificial and real light curves, as well as on already known CBPs are shown in Sect.~\ref{sec:test_results}. We present our results in Sect.~\ref{sec:results} and discuss the planet candidates that were found. In Sect.~\ref{sec:limits} we interpret the search results and derive upper limits for the probabilities that circumbinary planetary systems with detectable planets exist within the analyzed sample of binaries. Final conclusions are given in Sect.~\ref{sec:conclusion}.

\section{Issues on searches for transits of circumbinary planets}
\label{Sec:search}


The detection of transiting CBPs is quite different to that of planets orbiting single stars. In the case of a CBP, the configurations when transits occur depend on both the binary and the planet's orbital phases. Therefore, the transits occur only quasi-periodically, within a 'transit window' that is recurrent with the period of the planet. Furthermore, the shapes of transit light curve can be quite complex \citep[e.g.][]{deeg98}, depending on the relative velocities of the three bodies, and are further complicated by planet transits happening during mutual binary transits. Indeed, Kepler 16b and most of the first transiting CBPs were found using visual inspections, only some more recent and  more challenging discoveries \citep[e.g. Kepler-413b, ][]{kostov14}) were found by search algorithms. On the other hand, the complexity of these curves provides a secure diagnostics to assure the planetary nature of such a detection, freeing transiting CBP detections from the worries about false alarms that are notorious on single star transit detections \citep[e.g.][]{brown03,almenara09}. 

A first step in a CBP transit search is the removal of the binary eclipses. For this step, the binary eclipses may be modelled either by a physical model of the binary, which involves the fitting of physical binary parameters, or by a purely phenomenological description of the binary brightness variations, which may be derived from a light-curve that is phased with the binary's period. This second method is the one used in this work. In either case, the removal of the binary might be affected by period changes in the binary orbit, due to both evolutionary and dynamical effects which make it more difficult to totally remove the eclipse light curve. Non-eclipsing third components with stellar masses may also cause significant period variations. In the case of EB light-curves in Kepler data, in some cases such variations were on the order of per-cents of the period itself \citep{borkovits15}. In the case of the CBP search in CoRoT light-curves, we expect period variations to be however negligible. For one, the length of the light-curves is limited to $\approx$150 days, a time-span over which period variations will manifest themselves much less. For another, we are interested in a search of planetary tertiary masses, and such masses will not cause period variations that would cause a relevant effect onto the binary light-curve removal. In the known transiting CBP's from Kepler, the variations in period within 150 day time-spans were all below 1 minute. Furthermore, any strong period variations that might be introduced by stellar-mass tertiary would at worst cause us to miss a planet-detection in that particular system.

Besides the detection algorithm used in this work, outlined in more detail in Sect.~\ref{sec:method}, during the course of this work, two other detection algorithms of interest for the detection of CBPs became published:  
QATS \citep{2013ApJ...765..132C} 
searches for a most likely transit depth and duration that is subject to a quasi-periodic condition. That is, QATS is aimed at detecting events that have variations around a typical 'baseline' condition, as is found in transits across single stars that suffer variations in periodicity, depth, and duration, due to the influence from other orbiting planets. QATS was used for the detection of several planets with strong transit timing variations, but its ability for the detection of CBPs, whose transits  may have much more complicated deviations from a `baseline-transit' \citep[e.g.][]{deeg98}, is still to be demonstrated\footnote{The only application of QATS to CBPs we are aware of is a search for a further planet in the 2-planet CBP system Kepler 47, where QATS failed to detect the outer planet, being described by \citet[][in online supporting material]{2012Sci...337.1511O} as "very sensitive to detrending errors for longer periods".}.

Another detection method was described in the context of the discovery of Kepler 64b  \citep[][given there as KIC 4862625]{2013ApJ...770...52K}, 
which was the first discovery of a CBP from the employment of a transit detection algorithm\footnote{Simultaneously, its visual detection in Kepler light-curves was reported as `PH1b' by \citet{2013ApJ...768..127S}.}. 
Kostov's method is based on the widely used BLS algorithm \citep{2002A&A...391..369K}, 
but tuned to identify individual (instead of periodic) transit events as well as 'anti-transits', that is, brightening features with similar shapes.
A statistical comparison of transit and anti-transit events using the method of \citet{2006AJ....132..210B} 
leads to the identification of light-curves in which true transit events may be present, requiring then further visual inspection. However, we do not expect this method to perform well on Corot light-curves, since  systematical flux-variations are assumed to be free of strong tendencies towards either transit or anti-transit like events. In contrary to Kepler data, Corot data do however suffer from frequent flux-jumps that are mostly brightness increases (see Sect.~\ref{jump_correction}, thereby violating this assumption,  with residuals from the corrections of these jumps being neither symmetric in positive or negative flux. We expect therefore that at least some modifications would be required for that method to work well on CoRoT data.



\section{CoRoT binary sample}
\label{sec:bin_sample}

For our CBP search, EBs were selected based on the CoRoT N3 automatic variable star classification output tables \citep[][tables available in \url{http://idoc-corotn2-public.ias.u-psud.fr/jsp/CorotN3.jsp}]{debosscher09}. In order to not miss any potential CBP host system, we selected all variables classified as EBs with non-zero probability (3188 targets). This means that many of our targets are not binaries in fact. In order to get a more reliable statistic of CBPs we inspected all these light curves visually and sorted out the non-eclipsing binary targets, leaving 1512 light curves in the list.

Additionally we merged in a list of 1268 EBs identified with the DST transit detection tool \citep{cabrera12}, developed initially for the search of transiting planets in CoRoT data. These binaries have generally low eclipse amplitudes of <5$\%$ and have been published for some CoRoT runs (IRa01,  LRc01, SRc01, LRa01, LRa03 and SRa03) in \citet{carpano09, cabrera09, erikson12, carona12, cavarroc12}. Since many targets were observed in more than one CoRoT run, the combination of these two lists results in 2780 light curves of 2290 individual binaries (see Table \ref{tab:ID}), published fully as electronic version).

We note that a list of EBs that is included in an upcoming paper describing the full set of CoRoT's transit detections (Deleuil et al., in prep) has been derived independently to the list of EBs used in our work.

\begin{table}
\caption{Likely eclipsing binaries in CoRoT data, selected for the CBP search of this work. The period is the presumed binary orbital period from the procedure given in Sect.~\ref{sec:binperiod}. The full table with 2290 targets is available as electronic version.}
\label{tab:ID}
\centering
\begin{tabular}{cc}
\hline\hline
CoRoT-ID & Period \\
\hline
100552362 &  1.347850 \\
100588681 &  0.793206 \\
100619354 &  3.358470 \\
100624108 &  0.645742 \\
100657980 &  2.923230 \\
\hline\hline
\end{tabular}
\end{table}



\section{Detection method of CBP candidates in CoRoT data}
\label{sec:method}

Here we describe the method we developed for quasi-periodic transiting CBP search in the CoRoT EB light curves. The method is flexible enough to ingest light curves obtained by other telescopes (e.g. Kepler) as well, with only slight modifications in the parameters during the light curve preparation.

First we present the code itself and how it works. Afterwards we show tests regarding the capability of the code and apply it to the known CBP Kepler-35b \citep{welsh12}.

\subsection{Light curve preparation}

In order to search for transit-like signals, ideally all flux variations should be removed except the transit like features. These variations can be both of physical (eg. eclipses, stellar activity) and instrumental origin. In the case of CoRoT, sudden flux jumps are quite common in the light curves and longer-term trends may be present as well.

The main steps of the light curve preparation for each binary are as follows:
\begin{itemize}
 \item selection of valid data points;
 \item binning to 512\,s;
 \item linear trend removal;
 \item binary period refinement and stellar eclipse removal;
 \item jump correction;
 \item additional light curve corrections.
\end{itemize}

\subsubsection{Data point selection}

We use the CoRoT N2 data release of 2013, starting from the provided fits files. Only data points with $STATUS=0$ are selected, which means we don't use any data flagged as being infected by cosmic rays or taken when the satellite crossed the South Atlantic Anomaly.

\subsubsection{Data binning}

CoRoT data contain two different time sampling mode, 32\,s and 512\,s integration time, which could be alternated during the observations. In order to get a consistent light curve with lower uncertainty, we binned all 32\,s observations to 512\,s.

\subsubsection{Linear trend removal}
\label{sec:lintrend}
Due to several factors (aging of the CCD cameras; changing position of the sun relative to the satellite causing scattered light; gradual recovery from Cosmic-Ray hits that occurred in previous pointings), CoRoT light curves may have slight linear gradients, amounting to absolute flux variations of less than 2\% across individual CoRoT runs \footnote{\cite{2016arXiv160902436A} found flux variations of $\la 5\%$ across the entire 6-year long CoRoT mission, based on zero-point magnitudes from absolute photometry. There was however a significant increase in noise across the mission; e.g.  for an R=14mag star, photometric errors over time-scales of 2h incremented by a factor of $2.2$}. To calculate a more accurate average binary light curve it is useful remove such gradients. Furthermore, in several light curves there are large instrumental jumps which cause unusable binned binary light curves (see Sec.\ref{sec:binperiod}). In these cases a linear fit to the whole light curve smooths the distribution of the data points.

\subsubsection{Binary period refinement and stellar eclipse removal}
\label{sec:binperiod}
Initial binary orbital periods \citep[e.g. from the CoRoT N3 data product,][]{debosscher09} may not be of sufficient accuracy for the removal of the stellar eclipses from the light curves. Therefore, before we subtract the eclipses, we refine the orbital periods. This is done by trying different orbital periods around the catalogue value. For each test period we fold the light curve, divide it into bins, calculate the average in each bin, interpolate the binned points to the real data, subtract this from all points in the given bin, correct for instrumental jumps (see Sec.~\ref{jump_correction}) and calculate the scatter of the whole processed light curve. The number of bins is $N_{bin} = 2\times\sqrt{N_{data}}$, where $N_{data}$ is the total number of data points and $N_{bin} \geq 200$, otherwise the phase curve is not sufficiently well sampled and parts of the eclipses may remain in the processed light curve. The accepted orbital period and final processed light curve are those where the scatter of the residuals is minimal. After this step the average of the processed light curve is 0.

\subsubsection{Jump correction}
\label{jump_correction}

Jumps in the CoRoT light curves are well known features \citep{mislis10, aigrain09, mazeh09}. These sudden flux changes range from several percent down to a few tenth of a percent. Usually these jumps occur as sudden flux-increases that return gradually over days to weeks to previous values, although sudden resets to previous values have been observed as well. Also, in some cases, the flux drops and returns in a few hours to the original value, producing a transit-like flux variation. Since in most cases they are present only in one color \citep{mislis10}, they can be separated from real flux variations in brighter stars, which were observed in chromatic (multicolor) mode. But this is not possible for fainter targets observed in monochromatic mode.

After the stellar eclipse removal, in order to find the position of the jumps, we step through all data points and calculate the median of the previous and next 3 days. At a jump, the difference of the two medians will reach a local maximum. These maxima are the borders of the light curve sections, which have to be shifted to the same level. The length of the section before and after the given point is selected so as not to confuse transits as jumps. The individual sections were fitted with a third order polynomial, which was then removed from the light curve sections.

\subsubsection{Additional light curve corrections}

So far we have not treated flux variations caused by stellar activity, which increases with shorter binary periods. Since our targets are mostly close binaries, it is crucial to deal with this effect.

The method we apply is described by \citet{cabrera12} and uses a Savitzky-Golay filter. It eliminates long term light variations, like spots on the rotating surface of stars, but keeps short transients, like transit events, untouched. Variations within time scales of a few hours remain in the light curves, as these cannot be removed without degrading any transit signals as well.

After all these steps it is still possible to have some remnants of the eclipses in the processed light curves. This can be happen if the orbital period of the binary or the depth of the eclipses slightly changed during the observations. These remnants mimic periodic signals during the transit search and the real CBP transits may remain hidden. To avoid this, we fold the processed light curves with the binary orbital period, divide it into 200 bins and set all data points to zero in those bins, for which the following equation is true:

\begin{equation}
 \sigma_{bin,i} > \overline{\sigma_{bin}} + 3 \times \sigma_{\sigma_{bin,i}},
\end{equation}
where $\sigma_{bin,i}$ is the standard deviation of data points in the $i^{th}$ bin, $\overline{\sigma_{bin}}$ is the average of the standard deviation of the data points in the bins, while $\sigma_{\sigma_{bin,i}}$ is the standard deviation of the scatter in each bins ($\sigma_{bin,i}$).

\subsection{Quasi-periodic transit search}
\label{transit-search}

With the light curves treated as described, it is possible to search for transits in an automatic way.

\subsubsection{Planetary orbital period}
\label{planet-period}
The innermost stable orbit of a planet around a binary system is $\sim$ 3 times the orbital period of the binary \citep{dvorak89}. The actual value depends on the eccentricity of both the binary and the planet and the mutual inclination of the orbital planes \citep{holman99}. Therefore, we start the planet search with a conservative $P_{pl} = 2.6 P_{bin}$ and increment it up to half of $T_{cov}$, the length of the whole light curve. The steps are optimized in order to keep some overlapping between the searching phases incrementing from one period to the next one, even at the end of the observations. A minimal planetary orbital period of 2.0 days was also imposed. Binaries with correspondingly short periods (e.g. $P_{bin} < 0.77d $) are usually very active contact binaries. Planet detection attempts in the $P_{pl} < 2 $day regime led to multitudes of peaks in the search statistic $S_{total}$ (see Sect. \ref{sec:fin_stat}), which made it impossible to detect real planets in that regime.

\subsubsection{Transit parameters}

We search for simple box-shaped transits, which have only two parameters, their depth and duration. Contrary to the case of a single star, where all transits have the same depth and duration, in binary systems these parameters vary. The depth depends on the stellar component that is occulted, while the duration depends on the relative velocity of the transiting planet and the stars.

In duration we search for 1, 2, 3, 4, 6, 10 and 16 hour long transits. The transit depth is changed from $1.0 \times \sigma_{lc}$ up to 0.04 mag, where $\sigma_{lc}$ is the standard deviation of the processed light curve after the stellar eclipse removal, incrementing the depth in each step by a factor of $\sqrt{2}$. We also include a zero transit depth, since due to precession it may be possible that at some orbits no observable transits occur \citep[e.g.][on the issue of CBP's with sparse transits]{martin14}. These parameters are independent for each planetary test orbit.

\subsubsection{Search for quasi-periodic events}

We fold the processed light curve with the orbital test period of the planet. Then going through all phases we put a box-shaped transit at the same orbital phase for every planetary orbit. At each orbit we shift the test transit in phase individually in order to deal with the quasi-periodic nature of the transits. The shift interval depends on the ratio of the orbital period of the binary and the test period of planet, based on Kepler's third law with the planet of negligible mass, and it is within a few $\%$ of the planet orbit. The maximum shift in the epoch in units of fractional phase is:

\begin{equation}
 \mathrm{Shift}_{max} = \pm \frac{1}{2\pi} \times \arcsin{\frac{P_{bin}^{2/3}}{P_{pl}^{2/3}}}.
\end{equation}

This is an overestimation, since it assumes that all the mass of the central binary is concentrated in one component, while the true maximum elongation (of the components from the barycenter) is not the distance between the two stellar components, but the distance between the lower mass component and the barycenter. However, we don't take into account eccentric planetary or stellar orbits, which can increase the phase range of possible transits.

\subsubsection{Final statistic}
\label{sec:fin_stat}

After we try all the possible phase, shift, transit depth and duration combinations we save the best fit for each test period. The goodness of the fit for each set of parameters is given by the variance:

\begin{equation}
 S_{total} = \displaystyle\sum_{i=1}^{N} (F_{proc, i} - F_{test, i})^2, 
\end{equation}
where N is the number of data points in the light curve, $F_{proc, i}$ is the flux of the $i^{th}$ point in the processed light curve and $F_{test, i}$ is the flux of the $i^{th}$ point in the modelled light curve.

\subsubsection{Automatic detection}

Once we have the goodness of fit for all test periods we search for peaks in it. A sample case is shown in Fig. \ref{fig:kepler-35b}. If there is a periodic or quasi-periodic signal in the light curve, there are corresponding peaks in the period vs. $S_{total}$ diagram at the real period and its harmonics.

We detect peaks using the out-of-peak points to calculate a base value and the standard deviation.

\begin{equation}
 D_p = \frac{S_{total,\;p} - MEDIAN(S_{total,\;around\;p})}{\sigma_{S_{total,\;around\;p}}},
\label{eq_Dp} 
\end{equation}
where $D_p$ is the detection statistic for the $p^{th}$ period value

The peak is significant if the difference between the peak value and the median around the peak is 5$\sigma$ above the standard deviation of the points around the peak (D > 5.0). This detection threshold was chosen as it leads to a number of detections that is feasible for a subsequent manual revision (as described in Sect. 6.2), yielding a signal in $\sim$18\,\% of the binaries.

18\,\% is much above the expected frequency of transiting CBPs. Less than 1\,\% of the Kepler EBs have transiting planets and the expected number of detectable planets in the CoRoT light curves is even lower due to the lower photometric accuracy and the shorter length of its observations. But since the transit signals are week, we might miss the planets if we use a higher threshold. The 506 light curves in which a signal was detected were then revised individually (see Sect.\ref{sec:candidates}).


\section{Performance tests and detection probabilities}
\label{sec:test_results}


\subsection{Test planets in real CoRoT data}


In order to derive detection limits of the algorithm, we tested it on real CoRoT EB light curves with simulated transiting planets. Its results in terms of detection probabilities (e.g. the ratio of detected and all tests planets) versus relative transit depth and the number of transits in a lightcurve, $N_{tr}$, are summarized in Table \ref{tab:test_result}. They were derived by the following procedure:\\

\definecolor{gre}{rgb}{0.0, 1.0, 0.0}
\definecolor{yel}{rgb}{1.0, 1.0, 0.0}
\definecolor{red}{rgb}{1.0, 0.0, 0.0}

\begin{table*}
\caption{Detection probability (the ratio of detected and all tests planets) of the planet searching code on real CoRoT EB light curves. The total number of random tests performed for each box are in brackets. The accuracy of the probability values is  1 to 3 \%. Colors indicate ranges of detection probability: red <\,0.5, green >\,0.8, yellow 0.5-0.8.}
\label{tab:test_result}
\begin{center}
\small
\begin{tabular}{|l||l|l|l|l|l|l|l|l|l||l|}
\hline
$N_{tr}$ & \multicolumn{9}{c|}{$\Delta F$: Transit depth in RMS units} & Total \\ \cline{2-10}
 &  0.5 -- 1.0 &  1.0 -- 1.5 &  1.5 -- 2.0 &  2.0 -- 2.5 &  2.5 -- 3.0 &  3.0 -- 3.5 &  3.5 -- 4.0 &  4.0 -- 4.5 &  4.5 -- 5.0 & \\
\hline
\hline
2 -- 3 & \cellcolor{red}0.04 (501) & \cellcolor{red}0.18 (506) & \cellcolor{red}0.28 (537) & \cellcolor{red}0.38 (531) & \cellcolor{red}0.49 (520) & \cellcolor{yel}0.50 (505) & \cellcolor{yel}0.55 (519) & \cellcolor{yel}0.62 (470) & \cellcolor{yel}0.63 (527) & \cellcolor{red}0.41 \\
3 -- 4 & \cellcolor{red}0.04 (257) & \cellcolor{red}0.29 (253) & \cellcolor{red}0.46 (251) & \cellcolor{yel}0.63 (239) & \cellcolor{yel}0.62 (231) & \cellcolor{yel}0.71 (240) & \cellcolor{yel}0.71 (243) & \cellcolor{yel}0.73 (251) & \cellcolor{gre}0.81 (238) & \cellcolor{yel}0.55 \\
4 -- 5 & \cellcolor{red}0.10 (172) & \cellcolor{red}0.39 (173) & \cellcolor{red}0.47 (148) & \cellcolor{yel}0.73 (161) & \cellcolor{yel}0.76 (162) & \cellcolor{yel}0.77 (170) & \cellcolor{yel}0.77 (132) & \cellcolor{gre}0.82 (130) & \cellcolor{gre}0.92 (165) & \cellcolor{yel}0.63 \\
5 -- 6 & \cellcolor{red}0.16 (322) & \cellcolor{red}0.48 (310) & \cellcolor{yel}0.66 (285) & \cellcolor{yel}0.78 (258) & \cellcolor{gre}0.85 (298) & \cellcolor{gre}0.88 (321) & \cellcolor{gre}0.90 (301) & \cellcolor{gre}0.92 (308) & \cellcolor{gre}0.94 (328) & \cellcolor{yel}0.73 \\
6 -- 7 & \cellcolor{red}0.20 (168) & \cellcolor{yel}0.59 (203) & \cellcolor{yel}0.78 (185) & \cellcolor{gre}0.86 (212) & \cellcolor{gre}0.88 (199) & \cellcolor{gre}0.93 (189) & \cellcolor{gre}0.93 (178) & \cellcolor{gre}0.94 (189) & \cellcolor{gre}0.93 (206) & \cellcolor{gre}0.80 \\
7 -- 8 & \cellcolor{red}0.27 (348) & \cellcolor{yel}0.64 (332) & \cellcolor{yel}0.79 (321) & \cellcolor{gre}0.89 (318) & \cellcolor{gre}0.94 (313) & \cellcolor{gre}0.95 (352) & \cellcolor{gre}0.96 (347) & \cellcolor{gre}0.96 (297) & \cellcolor{gre}0.96 (324) & \cellcolor{gre}0.81 \\
8 -- 9 & \cellcolor{red}0.30 (262) & \cellcolor{yel}0.71 (258) & \cellcolor{gre}0.82 (236) & \cellcolor{gre}0.91 (236) & \cellcolor{gre}0.95 (229) & \cellcolor{gre}0.95 (229) & \cellcolor{gre}0.97 (240) & \cellcolor{gre}0.98 (239) & \cellcolor{gre}0.97 (247) & \cellcolor{gre}0.83 \\
9 -- 10& \cellcolor{red}0.38 (194) & \cellcolor{yel}0.71 (173) & \cellcolor{gre}0.89 (187) & \cellcolor{gre}0.95 (173) & \cellcolor{gre}0.95 (177) & \cellcolor{gre}0.96 (184) & \cellcolor{gre}0.99 (183) & \cellcolor{gre}0.96 (177) & \cellcolor{gre}0.99 (163) & \cellcolor{gre}0.86 \\
10 +   & \cellcolor{yel}0.55 (730) & \cellcolor{gre}0.88 (726) & \cellcolor{gre}0.94 (719) & \cellcolor{gre}0.96 (684) & \cellcolor{gre}0.98 (666) & \cellcolor{gre}0.98 (767) & \cellcolor{gre}0.98 (778) & \cellcolor{gre}0.99 (746) & \cellcolor{gre}0.99 (728) & \cellcolor{gre}0.92 \\
\hline
\end{tabular}
\end{center}
\end{table*}

Since every light curve is individual, the best way is to use all the binaries of the sample described in Sect.~\ref{sec:bin_sample}, instead of selecting some subsets. In each EB light curve we inserted a test planet with a random orbital period, distributed between $2.6 \times P_{bin}$ and half of the total data length. The transit depth $\Delta F$ was selected randomly between 0.5 and 5.0 times the scatter of the processed light curve, while the transit duration was chosen randomly between 1 and 16 hours for each individual transit. The test was repeated several times, in some cases focusing on different regions in the number of transits parameter ($N_{tr}$), in order to obtain similar numbers (on the order of a few hundreds) of tests in all of the cells of Table \ref{tab:test_result}, resulting a total number of 25.975 tests. This means an average of 9 different test planets were inserted in each CoRoT EB light curve. Then we checked whether there is a significant detection peak at the orbital period of the test planet, following the procedure of Sect.~\ref{transit-search}. The fraction of successful detections of all test planets in a given cell is presented in Table \ref{tab:test_result}.


Since the code is searching for quasi-periodic signals, it doesn't find monotransits. In curves with only 1 test transit, it finds the real monotransit and may find something else in the light curve.
With 2 transits in the light curve, the overall detection probability is $\sim40\%$, but for deep transits ($\Delta_F > 3$, where $\Delta_F = \delta F_{abs} /  \sigma_{lc}$, with $\delta F_{abs}$ being the absolute transit depth) it is above 80$\%$. For $\geq$3 transits we are able to detect almost all test planets, except the shallowest ones with a depth of only $0.5 \times \sigma_{lc}$. These are the limits of the code.

As can be seen in Table \ref{tab:test_result}, the code is able to find >50\% of the planets if the length of the observation ($T_{tot}$) is at least 3 times longer than the orbital period of the planet ($P_{pl}$), e.g. for $N_{tr} \geq 3$. The detection probability reaches 80\% at $N_{tr} \simeq T_{tot}/P_{pl} > 6$ for the whole tested parameter space. If shallow transits with $\Delta_F <1.0 $ are not taken into account, the 80\% limit is reached at $N_{tr} > 5$ and the detection probability goes up to 96\% for 10+ observed transits.

\subsection{Kepler-35b}

\begin{figure}[htp]
 \centering
 \includegraphics[width=0.45\textwidth]{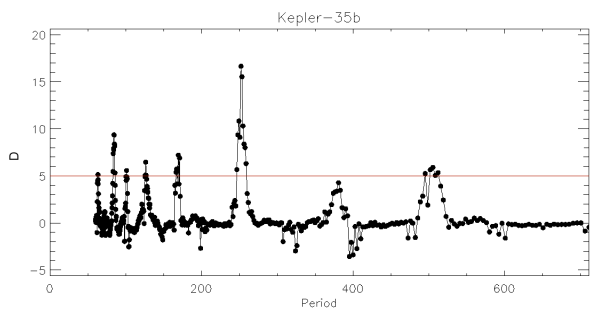}
 \caption{Planet search result for Kepler-35b. The highest peak corresponds to twice the real orbital period. However, the true period at $\sim$ 125 days is above the detection limit, too.}
 \label{fig:kepler-35b}
\end{figure}

As a further test, we attempted to identify the CBP Kepler-35b \citep{welsh12} from its light curve provided by the Third Revision of the Kepler Eclipsing Binary Catalog \citep{matijevic12}. The published orbital period of the planet is $\sim 131.5$ days. The rms of the processed light curve after the preparation is $0.05\%$, while the transit depth is $\sim0.3\%$, 6 times the rms. According to the tests in the previous section this should be an easy detection. The result of the automatic detection (planet orbital period vs. D) is shown in Fig. \ref{fig:kepler-35b}.


There are several peaks in Fig. \ref{fig:kepler-35b}, which correspond to the orbital period of the planet and its harmonics. The peak of the real period is at $\sim$ 125 days  ($5.2 \sigma$ detection), instead of the published 131.5 days. This is due to the relatively few transit events in the data with a large 'transit window' and the algorithm's design, which finds only approximate periods. The most significant peak at $\sim$ 250 days ($12.6 \sigma$ detection) corresponds to 2$\times$ the true period. This feature is not real, but a characteristics of the code itself. Since there is a high density of peaks in the $S$ statistics at shorter periods, peaks in the $D$ detection statistics become less significant, because the standard deviation in $S$ around a peak's period is larger (see Eq.~\ref {eq_Dp}). However, both peaks are above the detection limit of 5.0.


\section{Results}
\label{sec:results}

\subsection{Residual noises in the processed light curves}
\label{sec:residual}

The first part of the planet search was the removal of the EB light curves and of variations in flux-levels from cosmic rays or from instrumental effects. In an ideal case the such processed light curves contain only white noise and any planetary transits. However, it was impossible to remove all stellar activity variations as well as instrumental effects from the light curves -- especially in short period systems -- without destroying the transit signal.

\begin{figure}[htp]
 \centering
 \includegraphics[width=0.48\textwidth]{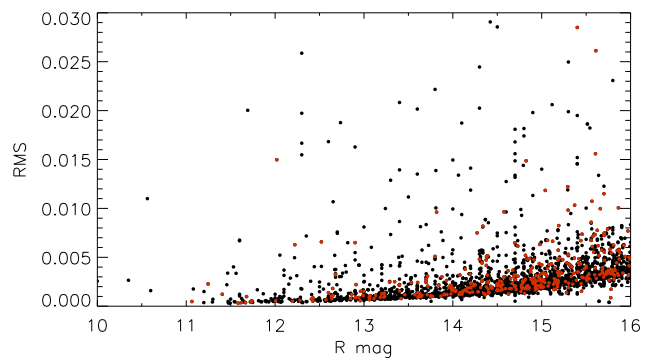}
 \caption{RMS ($\sigma_{LC}$) of the processed light curves after the removal of stellar eclipses and instrumental effects. Red dots represent the 506 targets with detections above the threshold.}
 \label{fig:R_rms}
\end{figure}

Fig. \ref{fig:R_rms} shows the RMS of the light curves after the eclipse variation is removed, as a function of the R band magnitude. Positive outliers are mainly due to stellar activity (e.g. short term fluctuations in the light curve), but incorrect jump removal may play a role as well. These light curves are not suitable for investigating stellar activity of binaries, due to the various corrections on the light curves in order to optimize them for transit search. However, variations on time-scales of transit durations cannot be corrected without affecting the transit signals.


\subsection{Planet candidates}
\label{sec:candidates}

\begin{figure}[htp]
 \centering
 \includegraphics[width=0.48\textwidth]{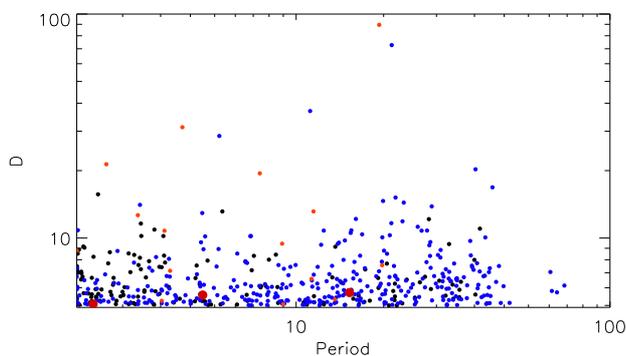}
 \caption{Detection statistic {\it D} vs. the planet orbital period in days (on a log-scale) in 506 light curves selected by the detection algorithm. A minimum search-period of 2 days was imposed, due to the reasons given in Sect.~\ref{planet-period}. The points' colours indicate classification from visual inspection. Red: planet candidates, orange: background binaries, black: stellar activity, blue: instrumental effect.}
 \label{fig:detection}
\end{figure}

During the automatic search we found 506 binary light curves with detections above the 5$\sigma$ limit (Red points in Fig.~\ref{fig:R_rms}).
We inspected these light curves visually and separated them in five groups based on the origin of the detected signal. In Fig.~\ref{fig:detection} we show the signal strength as a function of the detected period. The groups are as follows: background binaries (orange points, 21 systems), stellar variability (black, 143) and instrumental effects (blue, 339) and remaining planet candidates (red, 3).

The systems with the highest detection statistics belong to periodic variables, such as background EBs (e.g. there were two EBs in CoRoT's aperture
), spotted stars or pulsating variables.
In the long period regime, instrumental effects dominate, mainly due to smaller jumps still present in the light curves. These instrumental effect cannot be eliminated without modifying any real transit signals, as they are in the same timescale and amplitude range.

Several preliminary candidates were found, but all of them turned out to be mono-transits. Some of them were observed by CoRoT in chromatic mode, which means that their flux was measured in three different colour-bands. In these cases we checked the three light curves separately in order to recognize  events with strong colour dependencies. Unfortunately all of these candidates were false detections from instrumental effects.

\begin{table}
\caption{List of planet candidates. The last column contains the orbital period of the binary system.}
\label{tab:planet_candidate}
\begin{center}
\begin{tabular}{c|c|c}
\hline
\noalign{\smallskip}
CoRoT-ID & transit time & $P_{binary}$ \\
         & CHJD &  \\
\noalign{\smallskip}
\hline
\hline
\noalign{\smallskip}
310190466 & 3433.9 & 1.63353 \\
629951504 & 4222.9 & 0.26141 \\
634075176 & 4589.2 and 4591.9 & 5.02718 \\
\noalign{\smallskip}
\hline
\end{tabular}
\end{center}
\end{table}

\begin{figure*}[htp]
 \centering
 \includegraphics[width=0.96\textwidth]{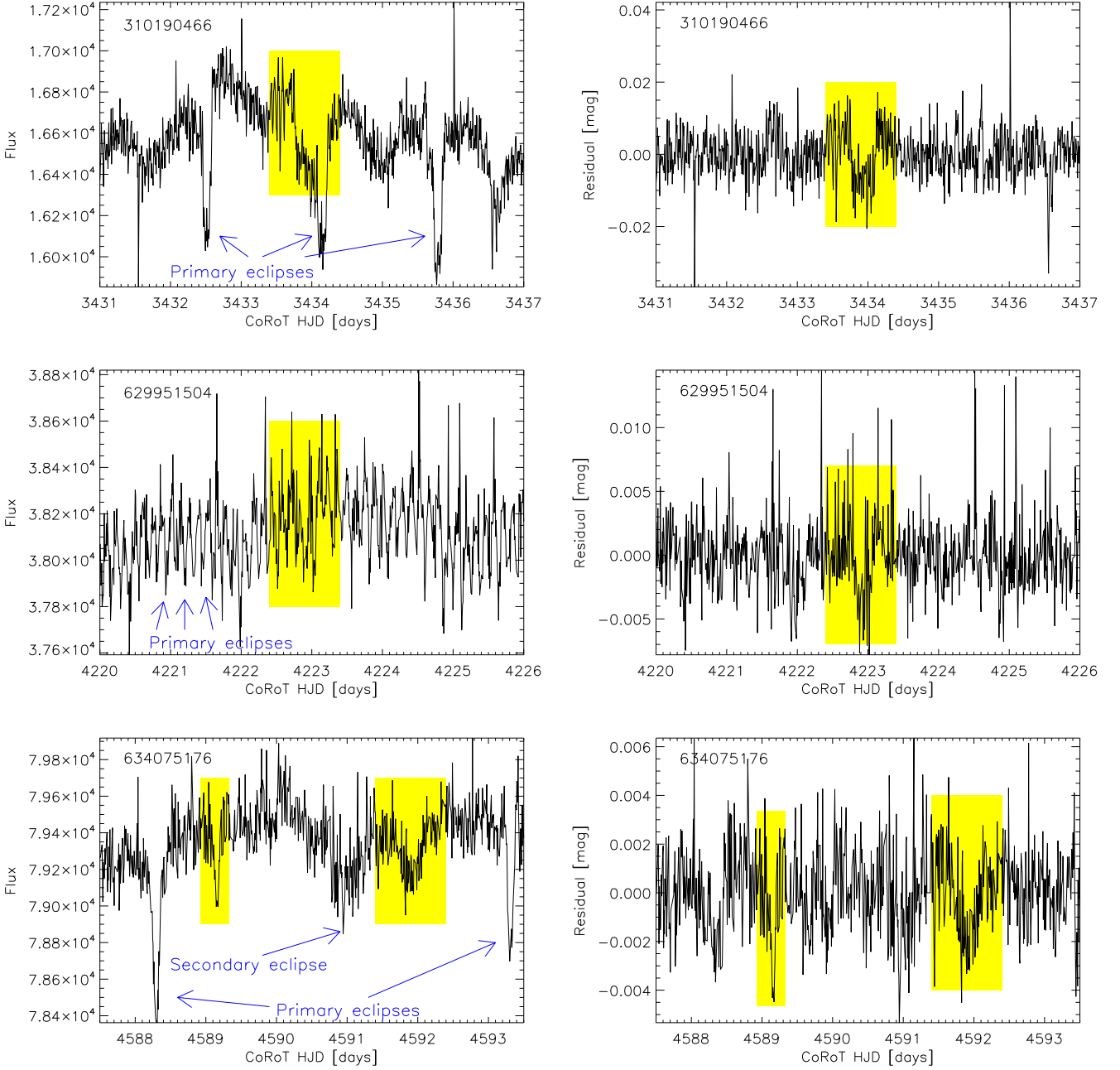}
 \caption{Light curves around potential transits (yellow background) of the three planet candidates. The original light curves are on the left side, with some stellar eclipses indicated, while the processed ones after the stellar eclipse removal are on the right side. Time is in CoRoT Heliocentric Julian Date, which is HJD - 2\,451\,545.}
 \label{fig:candidates_lc}
\end{figure*}

The candidates observed in monochromatic mode were checked individually, and only 3 of them remained for further investigation. These final planet candidates are listed in Table \ref{tab:planet_candidate}. In Fig. \ref{fig:candidates_lc} we present their original light curve and the processed one after the stellar eclipse removal around the suspected planetary transits.

CoRoT-310190466 is a relatively faint target with a brightness of R = 15.9 magnitude. Despite the large rms (0.0062 magnitude) there is a clear transit-like event at CHJD = 3433.9 (CHJD is the CoRoT Heliocentric Julian date, given by CHJD = HJD - 2\,451\,545.0). The depth of the suspected transit is 1.5\%, while the duration is $\sim$ 9 hours. The binary itself is an Algol-type, detached binary. The orbital period is 1.63353 days.

CoRoT-629951504 is a short period contact binary with an orbital period of 0.26141 days. The suspected transit is at 4222.9. The depth is 0.5\% and the duration is $\sim$ 7 hours. Since the transit shape could be very complex around a contact binary system, the small peak in the middle of the transit might be real, but most probably is only noise.

CoRoT-634075176 is a detached binary. The orbital period is 5.02718 days. There are two transit-like events in the light curve close to each other at 4589.2 and 4591.9. The depths are 0.4\%, while the durations are $\sim$ 3 hours and $\sim$ 9 hours, respectively. The residual feature at 4588.3 seems to be an instrumental jump. Folding the light curve of the binary, there is a shallow secondary eclipse at the orbital phase of 0.63, with an amplitude of $\sim$ 0.003, which implies stellar components of strongly different surface brightness. Therefore the transit candidates at CHJD = 4589.2 and 4591.9 are unlikely to be from a planet, since their similar depths would require stellar components of approximately equal surface brightness.

Note that the code is not optimized for finding single events. This means that mono-transits from longer period planets would be found only if there is some other transit-like event in the light curve (e.g. small jumps or eclipse residuals due to period change). Otherwise the monotransits remain hidden. In order to find these transits we checked all light curves visually searching for single transit-like events, but we did not find any additional candidates.

Regarding the large number of detections and high (100\%) fraction that turned out to be false alarms, which might leave doubts regarding the algorithm's efficiency, we note that many of these false alarms have been caused by features that are characteristic of CoRoT data. Similar situations may however also occur with data from the Kepler mission. When \citet{boyajian15} attempted an automatic search for further light-curves with features similar to the unusual object KIC\,8462852, they initially encountered over 1000 targets. A revision of these then led however to the discarding of all of them, with some of them being caused by instrumental effects. Similar to our search, their algorithm was also relatively broad in the requirements that led to a detection, being tuned to avoid the oversight of potential discoveries. As long as the revision of the detections that are encountered by such algorithms is manageable, be it manually or by specific algorithms, this is a correct approach.

\section{Limits to the occurence rate of circumbinary planets}
\label{sec:limits}

There are multiple works in the literature that establish abundances of some planet  populations from extrapolation of known abundances of somewhat different populations. For example, works about eta-Earth (abundance of Earth-like planets) extrapolate from known abundances of larger or hotter planets; e.g. \citet{2013PNAS..11019273P, 2014ApJ...795...64F, 2015ApJ...809....8B}. In this work, however, for one we are confronted with non-detections, and for another, the sample of known CBPs is very scarce and permits only very limited comparisons. We therefore attempt here to derive upper abundance-limits for given CBP-populations based on the absence of detections in our sample.

We calculate the upper limit of the potential abundance of a given planet in our EB sample by using the binomial distribution. The probability $P$ of getting exactly $k$ successes in $n$ trials is given by the probability mass function:
\begin{equation} \label{eq:binom}
 P = \binom{n}{k} p^{k} (1-p)^{n-k},
\end{equation}
where $p$ is the probability for the success of each individual trial. In our case $p$ is the unknown probability that a transiting CBP exists around an EB with a given set of parameters. In the case of a non-detection, we have $k=0$ and get then:
\begin{equation} \label{eq:binom0}
  P(k=0) = \binom{n}{0} p^{0} (1-p)^{n},
\end{equation}
which can be solved for $p$:
\begin{equation} \label{eq:binom1}
  p = 1 - P(k=0)^{1/n}.
 \end{equation} 
Since we are interested in approximate upper limits to the CBP planet abundance from non-detections, we set $P(k=0)$ to 0.5. This means that if the true fraction of planets with a given parameter-set is $p$, than there is 50\% chance that such planets would \emph{not} have been detected in the sample. The true value of $p$ from a non-detection remains however unknown and might be much smaller, so we can only determine reasonable upper limits $p_{max}$, given by:
\begin{equation} \label{eq:binom2}
  p_{max} \le  1 - P(k=0)^{1/n} = 1 -0.5^{1/n}.
 \end{equation} 
For more conservative values of $p_{max}$, the probability of getting no detection in $n$ trials, $P(k=0)$ should be set to lower values.

The suitability of a lightcurve in our EB sample to contain the potential detection of a CBP of a given period depends on the lightcurve's lenght and on the period of the EB.
In turn, the number $N_s$ of binary light-curves in the CoRoT sample that are suitable for the detection of a planet of a given period depends on that period. The light-curves' lengths varies from 23 days (SRa01 field) to 152 days (LRc01 field), and only light-curves longer than twice the orbital period of a potential planet were considered suitable, in order to assure that $N_{tr} \ge 2$. Furthermore, for a given planet period, only light-curves of binaries with periods shorter than 1/3.5 times the planet period are suitable, in order to consider only stable orbits.  The combination of these two requirements led to the values $N_s$ that are shown in Table~\ref{tab:planet_limits}, for different sets of binary periods.

Moreover, the detection probability of our code, $p_{det}$, is not 1.0, therefore an effective number $N_\mathrm{eff}$ of suitable systems for the detection of a planet with given { absolute transit depth $\delta F_{abs}$ and period $P_{pl}$ (resp. number of transits in the light curve, $N_{tr}$)} is being calculated:

\begin{equation} \label{eq:N_eff}
  N_\mathrm{eff} = \sum_{N_s} p_{det}(\Delta F, N_{tr}),
\end{equation}
with $p_{det}(\Delta F, N_{tr})$ being the values from Table 2, { where the relative transit depth $\Delta F $ and transit counts $N_{tr}$ have been calculated individually} for each of the light curves within $N_s$. For light curves for which a given test-planet results in relative transit depths of $\Delta F < 0.5$, $p_{det}$ was set to 0.

Inserting $N_\mathrm{eff}$ as $n$ into Eq. (7), we are able to obtain upper limits on the probability of existence of \emph{transiting} planets with $(P_{pl})$ and absolute transit depths $\delta F_{abs}$ in the CoRoT EB sample. 

In Table \ref{tab:planet_limits}, we give the maximum abundances for transiting Jupiter, Saturn and Neptune-sized CBPs, assuming $\delta F_{abs} = 0.5\%$ for the Jupiter, $\delta F_{abs} = 0.35\%$ for the Saturn and 0.05\% for the Neptune sized planets. These values for  $F_{abs}$ correspond to transit depths of planets with radii of 1 $R_{Jup}$, 0.8 $R_{Jup}$ and 0.3 $R_{Jup}$, respectively, across one component of a binary of two solar-like stars (with radii of $1 R_\sun$). 
We give upper limits for binaries with orbital period less than 1.0, 2.5, 5.0 and 10.0 days, while we are interested in planets up to 10, 25, and 50 days orbital periods. Note that the inner stability limit of the planets was set to 3.5 times the orbital period of the binary. This means that for $P_{bin} < 1.0$ day and $P_{pl} < 10.0$ days the upper { abundance} limit $p_{max}$ corresponds to planets between 3.5 and 10.0 days in period. Since we calculate upper limits, $N_{tr}$ was calculated using the longest planetary orbital period in the given period interval, which corresponds to the lowest $p_{det}$ values and consequently to the highest limits. For the upper limits that are based on $N_{eff} < 5$ (gray values in Table \ref{tab:planet_limits}), the CoRoT data do not place useful constraints on the planet abundances.

\begin{table*}
\caption{Upper limits for the abundance of eclipsing binaries with a transiting CBP.
}
\label{tab:planet_limits}
\begin{center}
\begin{tabular}{c||c|c|c|c||c|c|c|c||c|c|c|c|}
\hline
\noalign{\smallskip}
 & \multicolumn{4}{c||}{$P_{pl}$ < 10 days} & \multicolumn{4}{c||}{$P_{pl}$ < 25 days} & \multicolumn{4}{c|}{$P_{pl}$ < 50 days} \\
 & \multicolumn{4}{c||}{($i_{cr,pl} = 90.0^{\circ}\pm4.7^{\circ}$)} & \multicolumn{4}{c||}{($i_{cr,pl} = 90.0^{\circ}\pm2.5^{\circ}$)} & \multicolumn{4}{c|}{($i_{cr,pl} = 90.0^{\circ}\pm1.6^{\circ}$)} \\
Binary & $p_{max}$ & $N_s$ & $N_{eff}$ & actual $P_{pl}$   & $p_{max}$ & $N_s$ & $N_{eff}$ & actual $P_{pl}$   & $p_{max}$ & $N_s$ & $N_{eff}$  & actual $P_{pl}$  \\
& [\%] & &  & [days]  & [\%] & &  & [days]     & [\%] & &  & [days]     \\
 \noalign{\smallskip}
\hline
\hline
\noalign{\smallskip}
\multicolumn{13}{c}{Jupiter sized planet (transit depth = 0.005)} \\
\noalign{\smallskip}
\hline
$P_{bin}$ < 1.0 & 0.21  &  534 & 322.83 & 3.5 - 10.0  & 0.44 &  350 & 157.550 & 3.5 - 25.0  & 2.06 &  95 &  33.36 & 3.5 - 50.0 \\
$P_{bin}$ < 2.5 & 0.09  & 1151 & 748.60 & 8.75 - 10.0 & 0.17 &  774 & 407.911 & 8.75 - 25.0 & 0.56 & 318 & 123.24 & 8.75 - 50.0 \\
$P_{bin}$ < 5.0 &     - &    - &      - & -           & 0.11 & 1131 & 635.111 & 17.5 - 25.0 & 0.33 & 514 & 210.18 & 17.5 - 50.0 \\
$P_{bin}$ < 10.0&     - &    - &      - & -           & -    &    - &     -   & -           & 0.25 & 663 & 277.51 & 35.0 - 50.0  \\
\noalign{\smallskip}
\hline
\hline
\noalign{\smallskip}
\multicolumn{13}{c}{Saturn sized planet (transit depth = 0.0035)} \\
\noalign{\smallskip}
\hline
$P_{bin}$ < 1.0 &    0.46  &  534 & 149.52 &  3.5 - 10.0 &    1.33  &  350 &  51.58 &  3.5 - 25.0 & 6.27 &  95 &  10.71 &  3.5 - 50.0\\
$P_{bin}$ < 2.5 &    0.17  & 1151 & 407.35 & 8.75 - 10.0 &    0.41  &  774 & 170.14 & 8.75 - 25.0 & 1.38 & 318 &  49.76 & 8.75 - 50.0\\
$P_{bin}$ < 5.0 &        - &    - &      - &          -  &    0.24  & 1131 & 289.01 & 17.5 - 25.0 & 0.75 & 514 &  92.42 & 17.5 - 50.0\\
$P_{bin}$ < 10.0&        - &    - &      - &          -  &    -     &    - &      - &  -          & 0.56 & 663 & 122.54 & 35.0 - 50.0\\
\noalign{\smallskip}
\hline
\hline
\noalign{\smallskip}
\multicolumn{13}{c}{Neptune sized planet (transit depth = 0.0005)} \\
\noalign{\smallskip}
\hline
$P_{bin}$ < 1.0 &    7.75  &  534 &  8.59  &  3.5 - 10.0 & \textcolor{gray}{24.30}  &  \textcolor{gray}{350} &  \textcolor{gray}{2.49}   &  \textcolor{gray}{3.5 - 25.0} &    \textcolor{gray}{62.85} &  \textcolor{gray}{95} &  \textcolor{gray}{0.70} & \textcolor{gray}{3.5 - 50.0}\\
$P_{bin}$ < 2.5 &    1.50  & 1151 & 45.91  & 8.75 - 10.0 &  4.87  &  774 & 13.88   & 8.75 - 25.0 &   \textcolor{gray}{16.32} & \textcolor{gray}{318} &  \textcolor{gray}{3.89} & \textcolor{gray}{8.75 - 50.0}\\
$P_{bin}$ < 5.0 &        - &    - &     -  & -           &  2.57  & 1131 & 26.61   & 17.5 - 25.0 &     8.88 & 514 &  7.45 & 17.5 - 50.0\\
$P_{bin}$ < 10.0&        - &    - &     -  & -           &  -     &    - &     -   &  -          &     6.73 & 663 &  9.95 & 35.0 - 50.0\\
\noalign{\smallskip}
\hline
\end{tabular}
\end{center}
\end{table*}

In the case of 'misaligned' planets, e.g. those with a relevant angle between binary and planetary orbital planes, due to the precession of the orbital plane of the planet it is possible that transits occur only in a small fraction of planetary orbits. The time spent in transitability (when transits may occur due to correct inclinations; a detailed description of the geometry and transitability of misaligned CBPs is published by \citet{martin_tr15}) depends on the inclination of the binary, the mutual inclination of the planet and binary orbital planes
. However, if the misalignment of the planet is low enough, there will be a transit in every planetary orbit (e.g. 100\% transitability), regardless of the orbital phase of the binary. Therefore, the numbers of Table \ref{tab:planet_limits} are valid for planet inclinations of $90^{\circ} \pm 4.7^{\circ}$, $\pm 2.5^{\circ}$ and $\pm 1.6^{\circ}$ for orbital periods of 10, 25 and 50 days, respectively. {These $\pm$ values correspond to maximum misalignments that assure 100 $\%$ transitability around an edge-on ($i= 90^{\circ}$) binary with components of 1 solar radii. 
We note that the largest mutual inclination among the known transiting CBPs is 4.1$^{\circ}$ on Kepler 413b \citep{kostov14}. This planet has a period of 66 days around two smaller stars (of 0.78 and 0.48 $R_{sol}$) and hence displays year-long stretches without transits, on an 11 year precession period. The majority of known CBPs have however small mutual inclinations; e.g. \citet{kostov14} quotes an average of $0.3^{\circ}$ for the other CBPs known at that time, from which transit events at most of their planet orbits can be expected. }

The fraction of single stars with hot Jupiters (P < 10 days) is 1.2\% $\pm$ 0.38\% from the Doppler sample of \citet{wright12} and 0.6\% $\pm$ 0.1\% from the Kepler photometric sample \citep{wang15}. The difference is mainly due to different stellar types. In our binary sample the upper limit for CBPs of similar size is 0.21\% for $P_{bin} < 1.0$ days and 0.09\% for $P_{bin} < 2.5$ days. These results suggest that either there are much fewer short-period gas-giants (P < 10 days) in binary systems in nearly co-planar orbit than hot Jupiters around single stars, or most of such planets have highly misaligned orbits.

The smallest possible planet at 10 day period that has a potential to be found in any of the CoRoT binaries is $\sim 2.5 R_{Earth}$.



%

%

\section{Conclusions}
\label{sec:conclusion}

   \begin{enumerate}
      \item A catalogue of eclipsing binaries in CoRoT data was collected from the automatic classification results published in CoRoT N3 data and on from EBs that were identified with an algorithm used for the search for transiting planets around single stars. A total of 2780 EB systems were selected on which a search for circumbinary planets was performed.
      \item We developed a code for automatic transiting planet detection in EB light curves. Performance tests show that we are able to detect $>50\%$ of the test planets if the transit depth is $> 3 \times$ the rms noise of the processed light curve, after the removal of stellar eclipses and instrumental jumps, even if there are only 2 transits in the light curve. The detection probability is $>80\%$ if there are at least 5 transits.
      \item We have not found any planet candidates with transits from 2 or more orbits in the data. However, we have found 3 candidates that would have caused transits during one orbital revolution only. However, one of them (CoRoT-ID: 634075176), with an apparently double transit from the same orbital revolution, is unlikely to be a planet due to the transit's depths being incompatible with the stellar  components' surface-brightness ratio, which can be derived from the EB lightcurve.
      \item We calculated upper limits for the abundance of Jupiter, Saturn and Neptune sized planets in co-planar orbits for up to 10, 25 and 50 days orbital periods. Our results suggest that either there are much fewer short-period gas-giants (P < 10 days) in binary systems in nearly co-planar orbit than hot Jupiters around single stars, or most of such planets have highly misaligned orbits. The results are also valid for slightly misaligned planets, as long as such planets would generate transits at each planetary orbit.
      \item Due to the limited temporal coverage of the CoRoT pointings, the analyzed sample is mainly suitable for the search for short-periodic CBPs. The CBP with the shortest known orbital period is Kepler 47b, with a period of 49.5 days, which orbits a binary with a 7.45 day period. CoRoT would have been reasonably sensitive to a CBP with such an orbital period, but much less (or not at all) to any of the other, longer-periodic CBPs found by Kepler. In Kepler data, it has also been noted that most of the CBPS detected orbit near the inner stability limit, while none of the CBPs orbit binaries with periods of less than 7 days. Given that detection probabilities in Kepler data are also higher for the detection of short-periodic CBPs than for longer ones, the non-detection of CBPs in CoRoT data supports the pronounced absence of short-periodic CBPs. Our results give therefore support to the claim by \citet{hamers16}, that short-periodic binaries need a third stellar-mass companion for their formation, whose presence is however incompatible with the presence of larger planets near the inner stability limit.
    
   \end{enumerate}

\begin{acknowledgements}
      We thank the referee, Jon Jenkins, for comments that led to a significant improvement in the presentation of this paper. The authors wish to thank the entire CoRoT team which made the generation of the data used in this work possible. The CoRoT space mission was developed and operated by CNES, with  contributions from Austria, Belgium, Brazil, ESA, Germany and Spain. CoRoT data are available to the public from the CoRoT archive at http://idoc-corot.ias.u-psud.fr. PK and HD acknowledge support by grants AYA2012-39346-C02-02 and ESP2015-65712-C5-4-R of the Spanish Secretary of State for R\&D\&i (MINECO). This project has been supported by the Hungarian National Research, Development and Innovation Office -- NKFIH K-115709
      and by the Hungarian OTKA Grant K113117.
      We would like to thank Antonio Dorta for his help in the supercomputations.
      This paper made use of the IAC Supercomputing facility HTCondor (http://research.cs.wisc.edu/htcondor/), partly financed by the Ministry of Economy and Competitiveness with FEDER funds, code IACA13-3E-2493.
\end{acknowledgements}



\bibliographystyle{aa} 
\bibliography{cbp_ref} 

%
%
% ...
%
%
%
% ...
%
%

\end{document}